\documentclass[letterpaper,twocolumn,10pt]{article}
\usepackage{usenix,epsfig,endnotes}
\usepackage[ruled,linesnumbered]{algorithm2e}
\usepackage{tabularx}
\usepackage{siunitx}
\usepackage{enumitem}
\usepackage{booktabs}
\usepackage{graphicx}
\usepackage{subfig}
\usepackage{caption}
\usepackage{subcaption}

\begin{document}

\date{}

\title{\Large \bf LearnedKV: Integrating LSM and Learned Index for Superior Performance on Storage}

\author{
{\rm Wenlong Wang}\\
University of Minnesota, Twin Cities
\and
{\rm David Hung-Chang Du}\\
University of Minnesota, Twin Cities
}

\maketitle

\thispagestyle{empty}

\subsection*{Abstract}

We present LearnedKV, a novel tiered key-value store that seamlessly integrates a Log-Structured Merge (LSM) tree with a Learned Index to achieve superior read and write performance on storage systems. While existing approaches use learned indexes primarily as auxiliary components within LSM trees, LearnedKV employs a two-tier design where the LSM tree handles recent write operations while a separate Learned Index accelerates read performance. Our design includes a non-blocking conversion mechanism that efficiently transforms LSM data into a Learned Index during garbage collection, maintaining high performance without interrupting operations. LearnedKV dramatically reduces LSM size through this tiered approach, leading to significant performance gains in both reads and writes. Extensive evaluations across diverse workloads show that LearnedKV outperforms state-of-the-art LSM-based solutions by up to 4.32x for read operations and 1.43x for writes. The system demonstrates robust performance across different data distributions, access patterns, and storage media including both SSDs and HDDs.

\section{Introduction}
Unstructured data is projected to constitute over 80\% of all data collected globally by 2025 \cite{unstructured_data}, expanding at an annual rate of 55-65\%. Traditional relational databases struggle with such data's variability and complexity, leading to the rise of Key-Value (KV) stores \cite{Zhang2023PMDBAR, Cao2022ISHBaseAI, Atikoglu2012WorkloadAO, Eldakiky2021TransKVAN, Zhang2021NVLSMAP, Wu2020ACKeyAC, DeCandia2007DynamoAH, Cooper2010BenchmarkingCS, Rumbaugh2024TowardsSI}. KV stores organize data as key-value pairs, providing schema-free flexibility for diverse applications like caching, session storage, and large-scale data processing.

The Learned Index \cite{Kraska2017TheCF} has transformed indexing systems by learning key distribution patterns to predict data locations within a tolerable range. Unlike B+-Trees, it eliminates explicit mappings, reducing pointer traversals and index space while improving read performance. However, write/update and insert operations remain challenging as they can invalidate learned models by shifting key distributions. While updatable variants exist \cite{Wu2021UpdatableLI, Ding2019ALEXAU}, they struggle to match the performance of traditional indexes in larger-than-memory systems \cite{Lan2023UpdatableLI}.

Initially positioned as an in-memory index \cite{Kraska2017TheCF, Ferragina2019TheP, Kipf2020RadixSplineAS, Ding2019ALEXAU, Wu2021UpdatableLI, Li2021FINEdexAF, Tang2020XIndexAS, Galakatos2018FITingTreeAD}, the Learned Index faced significant challenges in persistence and recovery. System crashes would require complete index rebuilding through full data scans, making storage adaptation essential for practical deployments.

The Log-Structured Merge-Tree (LSM-Tree) \cite{ONeil1996TheLM} represents a crucial storage-centric indexing approach. It excels in write-intensive applications by organizing data into multiple layers and converting small random writes into larger sequential operations through periodic merging called compactions.  This design has spawned numerous variants \cite{RocksDB, LevelDB, Chen2021SpanDBAF, Kaiyrakhmet2019SLMDBSK, Wu2015LSMtrieAL, Kang2019TowardsBA, Lepers2019KVellTD, Wu2020ACKeyAC, Sarkar2020LetheAT}, with prominent implementations in production systems like Meta's RocksDB \cite{RocksDB}, Google's LevelDB \cite{LevelDB}, Amazon's DynamoDB \cite{dynamodb}, and Apache HBase \cite{hbase}.


Recent studies have explored LSM and Learned Index combinations \cite{Dai2020FromWT, AbuLibdeh2020LearnedIF, Lu2021TridentKVAR}. However, these approaches merely use the Learned Index as a supplementary path for SST files, overlooking its potential to replace traditional indexing mechanisms in a tiered structure that could fully leverage  the strengths of both designs.

We analyze key trade-offs between LSM and Learned Index structures: LSM provides efficient writes but suffers from increasing I/O amplification due to compactions and less read performance due to the required search in multiple layers as the  data grows, while Learned Index enables fast reads with compact space usage but requires sorted, preferably static datasets and struggles with updates. To leverage these complementary strengths, we propose a tiered architecture where LSM handles write operations and Learned Index accelerates read performance. While ideally static data would remain in the Learned Index and modifications would route to LSM, predicting update patterns is impractical. We therefore introduce a data migration mechanism that moves entries between the two indexes to optimize the overall system performance.

In many KV store applications, small keys are associated with significantly larger values, ranging from 100B to over 10KB \cite{Lu2016WiscKeySK, Ahn2016ForestDBAF, Cao2020CharacterizingMA, Atikoglu2012WorkloadAO, Debnath2010FlashStoreHT}. "KV Separation" \cite{Lu2016WiscKeySK} mitigates write amplification by storing complete key-value pairs in an append-only Value Log, while maintaining only keys and pointers in the LSM tree. For such systems, Garbage Collection (GC) helps maintain reasonable storage utilization by reclaiming space from outdated or deleted pairs. 

When storage space is constrained, GC triggers more frequently. During GC, we observed that in-memory key sorting incurs negligible overhead compared to data migration costs. This efficiency enables appending key-value pairs in sorted order, creating an ideal dataset for learned model construction. Since log entries remain static between GC cycles, the dataset provides the stability that is necessary for effective learned indexes. Consequently, converting the LSM to a Learned Index during GC becomes both feasible and efficient, significantly accelerating reads with minimal performance impact.

In this paper, we introduce LearnedKV, an efficient tiered key-value store that integrates LSM and Learned Index for superior storage performance. Through thoughtful design, this tiered architecture leverages LSM's high update throughput and Learned Index's read efficiency in a complementary manner.

Our key contributions are:
\begin{itemize}
\item We identify that GC-appended valid key-value pairs in KV separation designs provide an ideal dataset for Learned Index construction, addressing its storage-related limitations.
\item We propose LearnedKV, a novel tiered design where Learned Index directly references key-value pairs, reducing LSM size and I/O overhead while improving read performance.
\item We develop a non-blocking conversion mechanism that builds the Learned Index during GC while maintaining system availability.
\item Our evaluations show that LearnedKV outperforms existing LSM-based solutions by up to 332
\end{itemize}

The rest of this paper is organized as follows: Section 2 provides background, Section 3 presents our motivation and challenges, Section 4 details LearnedKV's design and implementation, Section 5 presents our experimental evaluation, Section 6 discusses future work and concludes the paper.

\section{Background}\label{sec:background}

\subsection{LSM and RocksDB}

The Log-Structured Merge-Tree (LSM-Tree) \cite{ONeil1996TheLM}, introduced in 1996, is widely adopted for persistent key-value storage. It efficiently transforms small random writes into large, sequential updates, beneficial for accessing SSDs and HDDs. And there are many researchers have explored methods to support large-scale key-value stores using the LSM structure \cite{Chen2021SpanDBAF, Kaiyrakhmet2019SLMDBSK, Wu2015LSMtrieAL, Kang2019TowardsBA, Lepers2019KVellTD}. RocksDB \cite{RocksDB}, developed by Meta, optimizes the LSM structure for fast storage media and write-intensive workloads.

\begin{figure}[ht]
  \centering
  \includegraphics[width=\linewidth]{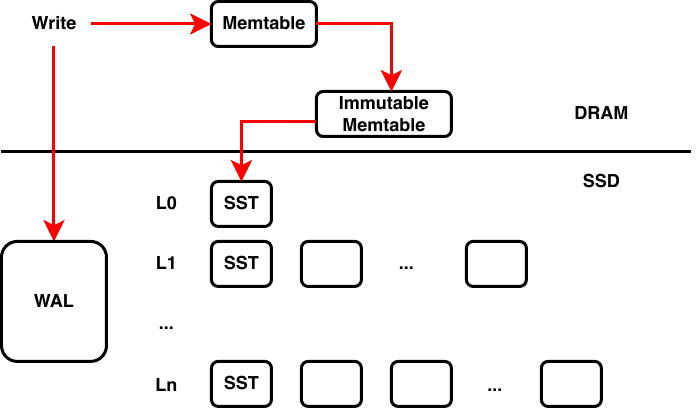}
  \caption{\textbf{RocksDB architecture.} In this figure and following sections, $L_0$ represents the highest level and $L_n$ represents the lowest level.}
  \label{fig:RocksDB}
\end{figure}

As shown in Figure \ref{fig:RocksDB}, RocksDB maintains in-memory Memtables and on-storage Write-Ahead Logs (WAL) and SST files. New writes enter a mutable Memtable, which, when full, becomes immutable and flushes to storage as an SST file. SST files are organized in levels ($L_0$ to $L_n$), with higher levels containing fewer files and recent updates and insertions. A background compaction process periodically merges files from a adjacent higher level to a lower level, removing redundant and  outdated entries.

Read operations search through Memtables and SST files across levels. While $L_0$ may require multiple file searches due to overlapping key ranges, lower levels guarantee at most one file search per level. RocksDB employs both file-level indexing and Bloom filters to minimize unnecessary I/O operations. However, the read performance is likely to be reduced with more data stored in more levels.


\subsection{KV Separation}
Real-world key-value workloads often feature small keys (8-16 bytes) paired with large values (100B to 10KB+) \cite{Lu2016WiscKeySK, Ahn2016ForestDBAF, Cao2020CharacterizingMA, Atikoglu2012WorkloadAO, Debnath2010FlashStoreHT}. KV separation techniques \cite{Lu2016WiscKeySK, Li2018EnablingEU, Tang2022FenceKVEE} address the resulting I/O overhead by storing values in a separate append-only log while keeping only keys and pointers in the LSM tree.

WiscKey \cite{Lu2016WiscKeySK} pioneered this approach, introducing garbage collection (GC) to reclaim space from outdated entries. HashKV \cite{Li2018EnablingEU} and FenceKV \cite{Tang2022FenceKVEE} further improved efficiency by partitioning the key space across multiple logs. The GC process is crucial for managing storage capacity, reducing space overhead, and maintaining system performance through efficient I/O management.

\subsection{In-memory Learned Index}
The Learned Index \cite{Kraska2017TheCF} introduced a novel approach to indexing by employing machine learning to predict key locations in sorted datasets. Initially developed for in-memory environments \cite{Kraska2017TheCF, Kipf2020RadixSplineAS, Wu2021UpdatableLI, Ding2019ALEXAU, Ferragina2019TheP, Li2021FINEdexAF, Tang2020XIndexAS, Galakatos2018FITingTreeAD}, it replaces explicit key-to-location mappings with lightweight models trained using methods like Linear Regression and Neural Networks.

Due to complex real-world key distributions, perfect predictions are likely unfeasible, requiring local searches around predicted locations to identify the target key. To bound worst-case performance, designs incorporate error bounds \cite{Kraska2017TheCF, Kipf2020RadixSplineAS} and hierarchical models that partition key ranges into smaller, more manageable segments. These approaches have demonstrated 1.8x to 3.2x read performance improvements over traditional indexes \cite{Kraska2017TheCF}.

\subsection{On-storage Learned Index with LSM}
As datasets grow beyond DRAM capacity, adapting Learned Indexes to SSDs has become crucial. While most of the pure Learned Index are yet to consistently outperform traditional B+-tree indexes \cite{Lan2023UpdatableLI}, some works are trying to integrate the Learned Indexes with LSM trees for large-scale key-value stores \cite{Wang2024LeaderKVIR, Liu2024LCKVLO, Dai2020FromWT, AbuLibdeh2020LearnedIF, Lu2021TridentKVAR}. Bourbon \cite{Dai2020FromWT} uses piecewise linear regression to optimize SST file lookups, while Google's approach \cite{AbuLibdeh2020LearnedIF} implements model-guided data placement. TridentKV \cite{Lu2021TridentKVAR} adapts model training based on workload patterns.

\begin{figure}[ht]
  \centering
  \includegraphics[width=\linewidth]{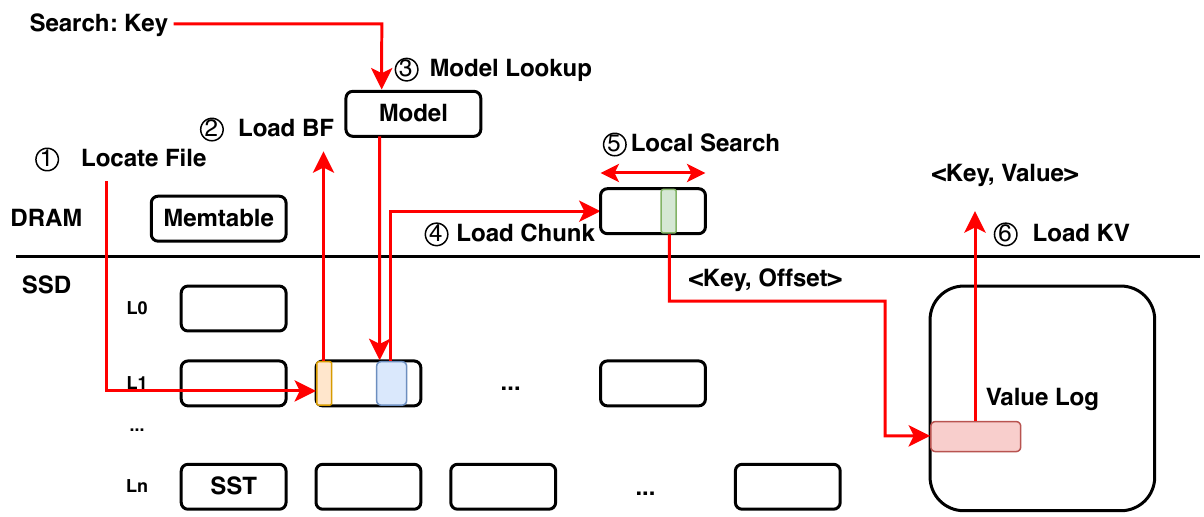}
  \caption{\textbf{Bourbon lookup process}}
  \label{fig:Bourbon lookup process}
\end{figure}

One representative of them is Bourbon \cite{Dai2020FromWT}, whose detailed architecture and lookup process are shown in Figure \ref{fig:Bourbon lookup process}. When searching for a target key, it firstly locates a candidate file and reads its Bloom Filter as traditional LevelDB/RocksDB does. If the Bloom Filter indicates the key's potential presence, it then queries the corresponding learned model of the SST to predict the location of the data chunk that should be loaded. Then it loads the data chunk and performs a local search to find the target key together with its offset in the Value Log. At last, it loads the target key-value pair from the Value Log. 

The Bourbon, together with some other Learned Index-LSM combination, primarily treat the Learned Index as a supplementary component for SST file optimization rather than fundamentally redesigning the index structure. As shown in Figure \ref{fig:Bourbon lookup process}, Bourbon still relies on LSM procedures for file location while using learned models only for intra-file searches. This pattern of limited integration is common across existing solutions, suggesting opportunities for more comprehensive architectural changes.

\section{Motivations and Challenges}


\subsection{Reducing LSM Size with Learned Index for Enhanced Performance}


Despite there are some research work regarding the on-storage Learned Index explored recently \cite{Dai2020FromWT, AbuLibdeh2020LearnedIF, Lu2021TridentKVAR}, 
there are still several limitations inherent to these designs, which guide us toward our proposed architecture: (1) Current approaches typically bind Learned Index models to specific SST files, necessitating frequent rebuilds during LSM compaction and resulting in redundant I/O operations. (2) Moreover, these designs often underutilize the Learned Index's potential as a standalone, compact, and efficient indexing structure, with its size being up to two orders of magnitude smaller than traditional indexes when indexing the same amount of data \cite{Kraska2017TheCF}.

As data volumes expand, LSMs must manage increasingly large-scale data within acceptable latencies. LSM storage usage often grows faster than the rate of raw key-value pair increase because there may be multiple versions of the same key within different levels. Despite compaction mechanisms and level additions, LSM size inevitably expands, impacting both read and write performance:
\textit{Read performance} degrades with increasing tree depth, as worst-case lookups may traverse to the bottom level, checking key availability at each level.
\textit{Write performance} suffers from increased compaction frequency and cost due to more SST files, affecting write throughput as compaction overhead is amortized across operations.

\begin{figure}[ht]
  \centering
  \includegraphics[width=\linewidth]{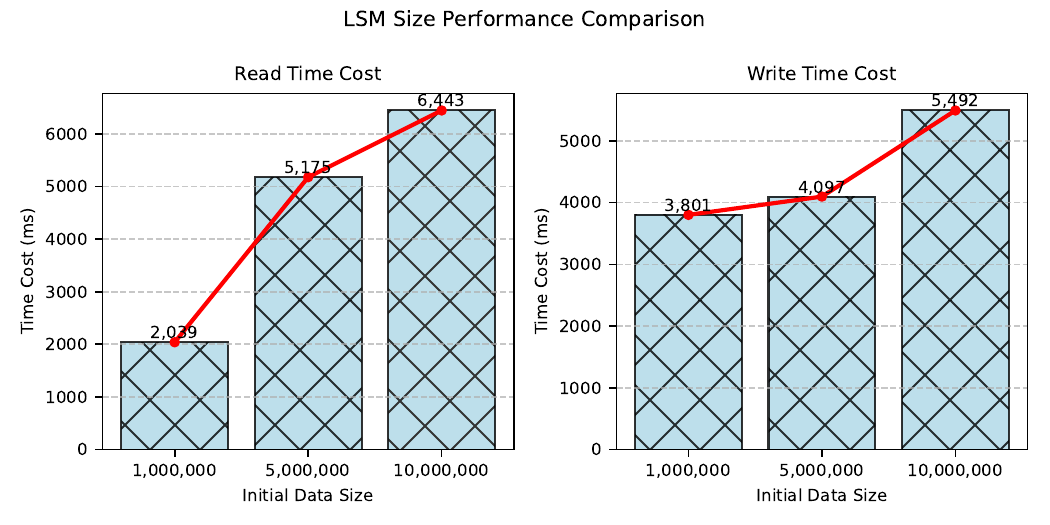}
  \caption{\textbf{Read and write time cost of the RocksDB across different stages}}
  \label{fig:LSM_size_impact}
\end{figure}

To quantify these effects, we conducted experiments using RocksDB \cite{RocksDB} with initial data sizes of 1, 5, and 10 million keys, measuring time costs for 1 million read and write operations each. In the experiment, we won't trigger any GC but will enable background compaction as configured by the default settings. Figure \ref{fig:LSM_size_impact} illustrates the performance degradation as LSM size increases, with read time cost rising by 216\% and write time cost increasing by 44.5\% when the dataset size increases from 1 to 10 million keys.

While existing studies on accelerating LSMs with on-storage Learned Indexes \cite{Dai2020FromWT, AbuLibdeh2020LearnedIF, Lu2021TridentKVAR} have focused primarily on replacing SST file index blocks, they overlook opportunities for LSM size reduction. This gap motivates our tiered design that leverages Learned Indexes to reduce LSM size, potentially improving both write throughput (through smaller LSMs) and read throughput (via optimized Learned Indexes), especially for larger datasets.

\subsection{Leveraging Complementary Strengths of Learned Index and LSM}

As discussed in Section \ref{sec:background}, Learned Indexes demonstrate superior read performance over traditional B+-Trees, showing 1.8x to 3.2x improvements for static in-memory datasets \cite{Kraska2017TheCF}. However, their performance in SSD environments and update efficiency remain limited \cite{Lan2023UpdatableLI}. Conversely, LSMs excel in handling update-intensive workloads on SSDs due to their ability to transform small random writes into large sequential operations \cite{RocksDB}. This complementary nature motivates our hybrid design that combines Learned Indexes' fast lookups with LSMs' efficient update handling.

Unlike existing solutions \cite{Dai2020FromWT, AbuLibdeh2020LearnedIF, Lu2021TridentKVAR}, we propose a tiered index where LSM absorbs random writes while Learned Index accelerates lookups. This design offers several advantages:

\begin{itemize}
\item \textit{Lookup performance}: Our system queries LSM first, then Learned Index if necessary, or accesses both simultaneously in multi-threaded scenarios. Based on Bourbon's "level learning" concept \cite{Dai2020FromWT}, the Learned Index can speed up read queries by up to 92\%. The overhead from LSM failure lookups remains minimal due to reduced LSM size.
\item \textit{Reduced LSM size}: Learned Index's partial dataset indexing reduces LSM size and compaction costs. Additionally, building the Learned Index during GC eliminates post-GC re-insertion overhead and cache pollution.
\item \textit{Write efficiency}: Support writes with pure LSM-level latency through Memtable handling.
\end{itemize}

The elegance of our approach lies in its simplicity and effectiveness - by fully decoupling these components, each indexing structure can operate in its optimal domain. Therefore, this design can potentially outperform current state-of-the-art solutions in both read and write operations while maintaining robust worst-case performance.

\subsection{Read-Optimized Indexing for Post-GC Static Data}
\begin{figure}[ht]
  \centering
  \includegraphics[width=\linewidth]{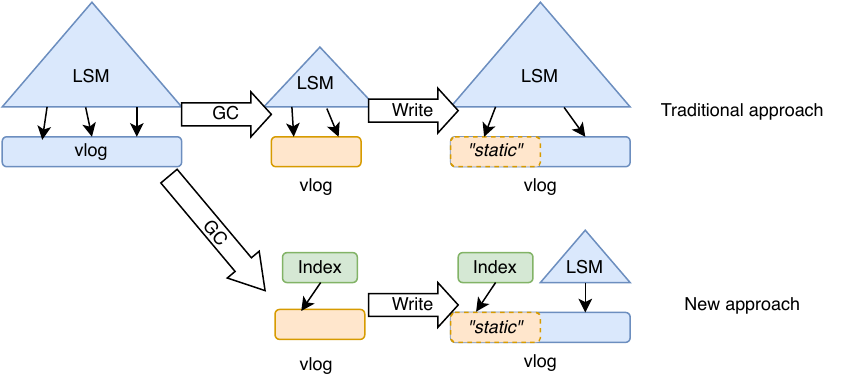}
  \caption{\textbf{Using read-optimized index (Learned Index) to index the "static" data}}
  \label{fig:post_GC_indexing}
\end{figure}
As shown in the upper part of Figure \ref{fig:post_GC_indexing}, traditional KV-separated LSM systems periodically trigger GC to compact both the LSM and vlog by removing duplicates. Then, subsequent writes append new key-value pairs to the vlog after the compacted section (yellow part in the figure), creating a "static" region that only serves reads without further modifications. We call this region "static" because after GC processes the data, these key-value pairs remain unchanged until the next GC cycle—they are not updated in place but instead, any modifications result in new entries appended at the end of the vlog. This stability makes the region ideal for read-focused optimization since its data distribution and location remain fixed. This observation reveals a significant opportunity: the post-GC static data could benefit from a different indexing structure than LSM.

LSM trees excel at write performance through buffering and compaction but introduce overhead for read operations, particularly as tree depth increases. For static data that no longer changes, this write optimization becomes unnecessary overhead. In contrast, specialized read-optimized indexes like Learned Indexes can provide significantly faster lookups when operating on stable data distributions. The top portion of Figure \ref{fig:post_GC_indexing} illustrates the traditional approach where LSM continues to index all data, including the static region. The bottom portion shows our proposed approach, where we strategically apply a read-optimized index (specifically a Learned Index) to the static data while maintaining LSM only for the active portion that receives new writes.

This hybrid approach capitalizes on the complementary strengths of both indexing structures: LSM efficiently handles write-intensive workloads with its batching and compaction mechanisms, while the Learned Index provides superior read performance for the stable data patterns in the static region. By applying the appropriate index to each data region based on its access pattern, we can optimize both read and write performance without compromising either.

\subsection{Challenges}

The proposed tiered index architecture presents several significant technical challenges. As the LSM component continuously absorbs random writes while the Learned Index remains static, the Learned Index gradually becomes outdated, diminishing its performance advantages over time. This necessitates an efficient mechanism to periodically migrate data from the LSM to the Learned Index while minimizing overhead. 

Furthermore, a robust KV store must maintain uninterrupted operation during internal structure modifications, requiring a conversion mechanism that preserves acceptable request latency without sacrificing data consistency or durability. With data distributed across two indexing structures, efficient query routing becomes essential—the system must efficiently determine which index to query first and how to handle potential duplication when the same key exists in both indexes. 

These challenges highlight the complexity of effectively integrating two fundamentally different indexing paradigms. In the following section, we describe how LearnedKV addresses these challenges through its novel architecture and conversion mechanisms that efficiently transform LSM data into the Learned Index during garbage collection processes while maintaining continuous system operation.

\section{LearnedKV}\label{sec:LearnedKV}

\subsection{Architecture and Basic Operations}





\begin{figure}[ht]
    \centering
    \subfloat[\textbf{Write operation in LearnedKV}]{%
        \includegraphics[width=0.9\linewidth]{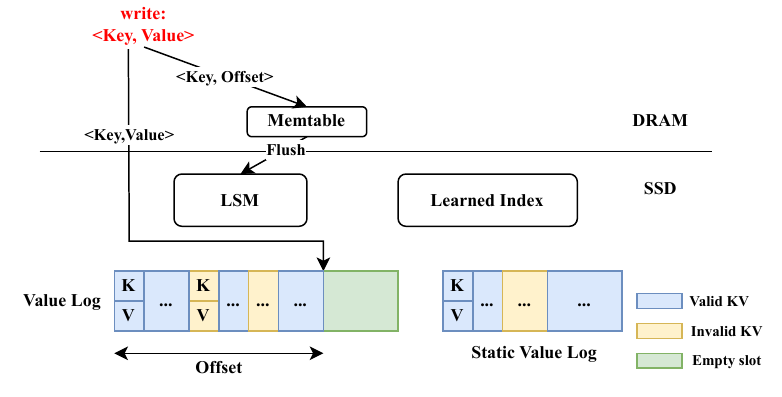}%
    }\\
    \subfloat[\textbf{Read operation in LearnedKV}]{%
        \includegraphics[width=0.9\linewidth]{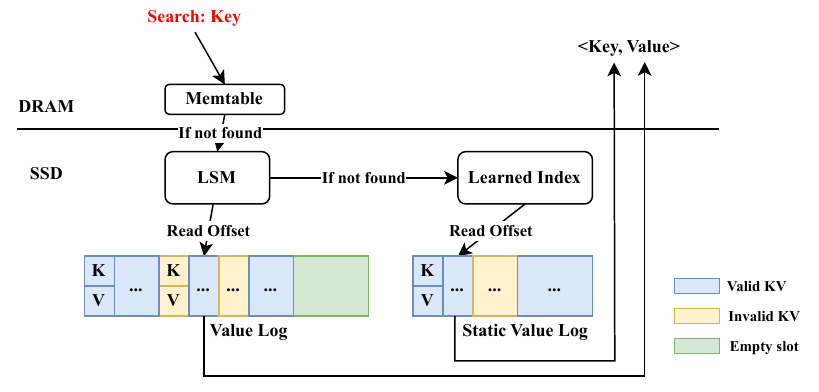}%
    }
    \caption{\textbf{LearnedKV Architecture}}
    \label{fig:architecture}
\end{figure}

We propose LearnedKV, an efficient tiered key-value store combining an LSM and a Learned Index for high-performance read and write operations on SSDs. Figure \ref{fig:architecture} illustrates the overall architecture and basic operations.

LearnedKV comprises three main components: an LSM Tree, a Learned Index, and two Value Logs. The "Value Log" is append-only and indexed by the LSM, while the "Static Value Log" stores key-value pairs indexed by the Learned Index and remains static until a GC process occurs. For clarification, unless specified, the "Value Log" will be referred to as the one indexed by the LSM in our later discussion.

The Value Log, following common practice, is fixed-size and accumulates both valid and invalid key-value (KV) pairs until a GC process is triggered. Valid KV pairs are newly inserted or never updated entries. When a new KV pair with an existing key is appended, the old entry becomes logically "invalid". Both LSM and Learned Index store key-offset pairs, where the offset indicates the KV pair's location in the respective Value Log.

During a write operation, the key-value pair is first appended to the Value Log, and its position is recorded as the "offset". The <$key, offset$> pair is then inserted into the Memtable. When the Memtable reaches capacity, it is flushed to storage and compacted into the LSM tree, following a process similar to RocksDB.

\begin{figure}[ht]
  \centering
  \includegraphics[width=\linewidth]{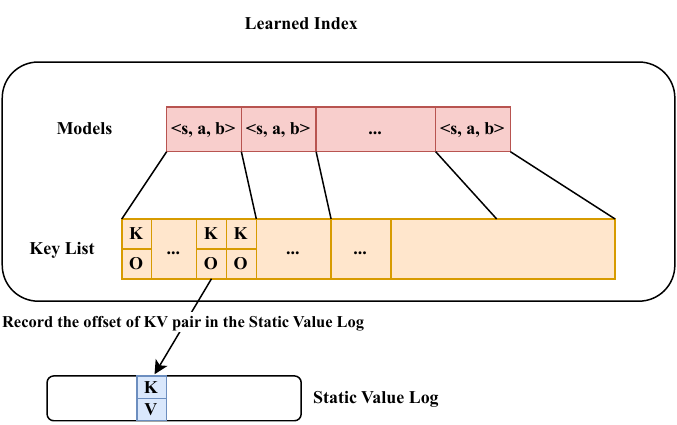}
  \caption{\textbf{Learned Index Architecture}}
  \label{fig:learned_index_arch}
\end{figure}

For point search requests, we first query the in-memory Memtable, returning the result immediately if found. If not, we search the on-storage LSM, followed by the Learned Index if necessary. Figure \ref{fig:learned_index_arch} illustrates the Learned Index structure, comprising a list of models and a "Key List" file.
Following the trend of prior studies, we employ Linear Models in our Learned Index. Each model is defined as a starting key (s), slope (a), and intercept (b). To search, we use binary search to find the largest starting key (s\_i) not exceeding the target key, then apply the linear model y = a*s\_i + b to predict the offset in the "Key List" file. This file contains sorted Key-Offset pairs, where "Offset" indicates the KV pair's position in the Static Value Log.
The predicted Key-Offset pair and surrounding pairs within the error range are loaded into memory. Once identified, the offset is used to retrieve the full KV pair from the Static Value Log. Similarly, if a key is found in the LSM, its offset is used to load the KV pair from the Value Log.
When a key exists in both the LSM and Learned Index, we prioritize the LSM version, as our mechanism ensures LSM-resident keys are more recent and thus valid.

It's worth noting that while our design significantly reduces LSM size—often to a point where it could theoretically fit entirely in memory—we deliberately maintain the LSM tree on persistent storage rather than keeping it exclusively in DRAM. This design choice ensures durability and crash recovery capabilities, as the LSM component contains the most recent updates to the key-value store. This persistence strategy maintains the robustness of traditional LSM-based systems while leveraging our tiered approach for superior performance, striking an optimal balance between durability and efficiency.

\subsection{GC-triggered Conversion from LSM to Learned Index} \label{sec:conversion}

A key contribution of our work is the ability to efficiently convert the entire LSM into a Learned Index during garbage collection (GC) while maintaining uninterrupted system operation. This process addresses the challenge of building a Learned Index, which typically requires a sorted, static dataset while ensuring continuous service to user requests.

\begin{figure*}[htbp]
  \centering
  \includegraphics[width=\textwidth,height=0.18\textheight]{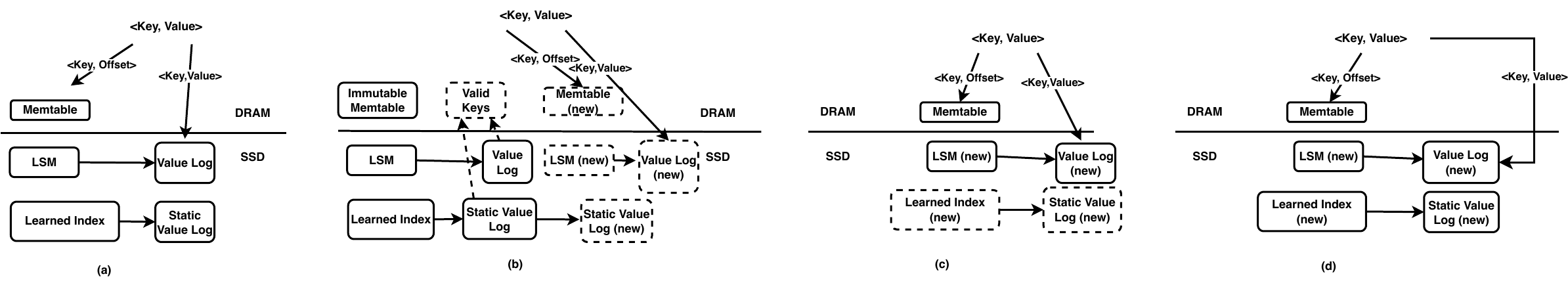}
  \caption{\textbf{Non-blocking Garbage Collection and Conversion.} \textit{The dashed box means the file is newly created in the current step.}}
  \label{fig:Conversion}
\end{figure*}

As shown in Figure \ref{fig:Conversion}, our GC and conversion process, inspired by HashKV \cite{Li2018EnablingEU} and FenceKV \cite{Tang2022FenceKVEE}, operates in four phases. Initially, the system runs normally using LSM, Learned Index, Value Logs on SSD, and an active Memtable. Upon GC initiation, we create a new LSM tree and Value Log for future writes, freeze the old Memtable, and collect valid key-value pairs from the old Value Log. We then sort these valid pairs in memory ($O(nlogn)$ time complexity), migrate them to a new Static Value Log, and construct a new Learned Index. This in-memory sorting incurs negligible costs compared to storage access latencies, requiring less than 80MB of temporary memory for our 10M-entry benchmark, which is released immediately after construction, making it feasible even on memory-constrained systems.

Our non-blocking GC process is designed to handle complete dataset compaction without imposing strict time constraints. Unlike traditional approaches where GC must complete quickly to minimize service disruptions, LearnedKV's architecture supports concurrent user operations throughout the entire GC cycle. By creating separate data structures for incoming writes while processing the old ones, we enable continuous service without performance degradation. This approach allows thorough processing of the entire dataset for optimal space reclamation and index construction without time pressure, leading to better storage utilization and more efficient indexing. This capability represents a significant advantage over systems that must compromise between GC completeness and operation latency.

To be noticed that, only the <$key, offset$> pairs will be stored and sorted in the memory, the large-sized key-value pairs are directly rewritten to the new static log. This approach is memory-efficient, requiring only megabytes of memory since the key-offset pairs are typically much smaller than the full key-value pairs. The Key List layer maintains logical sorting of key-value pairs while allowing physical storage to remain optimized for I/O performance.

This approach ensures continuous operation during GC while efficiently managing data and index structures. Write operations remain uninterrupted due to the separate new LSM and Value Log. For reads during GC, we query both old and new structures, with performance impact mitigated by the Learned Index's optimization for read operations.
The sorted valid key-value pairs in the new Static Value Log, along with a "Key List" file containing sorted keys and their locations, provide an ideal static, sorted dataset for building the Learned Index. This file is recreated during each GC cycle and remains unmodified between cycles. The resulting Learned Index, containing all valid keys from the LSM in a more compact and read-efficient format, replaces the old LSM tree. We reclaim the space occupied by the old LSM and value log, creating a new LSM for incoming writes.

This combined GC and conversion process effectively addresses both the challenge of efficient data conversion from LSM to Learned Index and the need for non-blocking support during conversion, ensuring system responsiveness and performance optimization throughout the process.

\subsection{Learned Index: Greedy-PLR+} \label{sec:learned_index}


While our design can accommodate various Learned Index models (e.g., RS\cite{Kipf2020RadixSplineAS}, RMI\cite{Kraska2017TheCF}, ALEX\cite{Ding2019ALEXAU}), we prioritize minimal overhead in model building and querying, without requiring in-place update capabilities. Following Bourbon \cite{Dai2020FromWT}, we adopt Greedy-PLR (Piece-wise Linear Representation) \cite{Xie2014MaximumEP} as our baseline statistical model and leave the exploration of the best-fit model for future research.

Our variant, Greedy-PLR+, focuses on page distance rather than absolute location distance, better suiting block device characteristics. Greedy-PLR+ builds piece-wise linear segments from sorted key-location pairs with $O(n)$ complexity. Starting with an initial segment from the first two points, it progressively adds points within a defined page-number error bound or creates new segments when the bound is exceeded. This produces a compact model list, each entry containing (start\_key, slope, intercept).

Our query processing uses two files: a sorted "Key List" containing (key, offset) pairs and cached "Models" in memory storing line segments. Given the Models' extremely compact size - our evaluation (Table \ref{tab:size_comparison}) shows it occupies only kilobytes, about 1/1000th the size of other components - the memory overhead is negligible. Queries first search the in-memory Models to locate the appropriate segment, predict the key's position, and load a bounded data chunk from the Key List. The matched offset is then used to retrieve the key-value pair from the Value Log, minimizing overall I/O operations.

\subsection{Range Scan}


Range scan support is essential for modern, large-scale KV stores. LearnedKV's tiered architecture necessitates efficient scanning across both the LSM tree and Learned Index to ensure comprehensive data retrieval. Our LSM implementation, RocksDB \cite{RocksDB}, provides a built-in iterator API for range scans, enabling sorted sequential access to key-value pairs within a specified range. RocksDB maintains individual iterators for each Memtable and SST file, managed by a "MergingIterator" that exposes them as a sorted stream.


For the Learned Index, keys are stored in the sorted "Key List" file. During a range scan, LearnedKV predicts the locations of start\_key and end\_key using the models, loads the surrounding data chunks, and determines their exact positions. Intermediate chunks are then loaded to identify keys within the scan range.

To merge the sorted key lists from the LSM and Learned Index, we iterate through both, adding the smaller key to the result list. For duplicate keys, we prioritize the LSM version as it contains the most recent updates. This process continues until both lists are exhausted. Finally, we retrieve the corresponding key-value pairs from the Value Log/Static Value Log within the specified range and return them to users.

Our experiments in Section \ref{sec:overall performance} show that LearnedKV's range scan operations outperform pure RocksDB \cite{RocksDB} by up to 4x for the same key range. This improvement can be contributed from two key factors:

First, LearnedKV reduces the time required for level-by-level key searches. While RocksDB's iterator presents keys in sorted order, they remain physically distributed across different levels and files. Maintaining a "MergingIterator" across these disparate locations incurs significant overhead compared to the Learned Index's direct, consecutive data chunk reads.

Second, key-value pairs in the Learned Index are stored contiguously in the Value Log. During the GC process (Section \ref{sec:conversion}), valid key-value pairs are appended to the Value Log in sorted order and used to construct the Learned Index. This sorted structure in both the Key List and Value Log remains intact until the next Learned Index construction, enabling efficient range scans. Conversely, the LSM's Value Log appends key-value pairs based on request timing, not key order, potentially scattering related pairs across the file and leading to less efficient random reads and page fragmentation.

\subsection{Optimizations}
While our general design is effective, we propose additional optimizations to enhance performance in specific scenarios. These optimizations are not essential concepts of our design but can offer benefits in certain circumstances.

\subsubsection{Range-query Assisted Conversion}

Our standard approach triggers conversion during GC, incurring minimal extra cost for building the Learned Index and avoiding LSM re-insertion. However, in over-provisioned storage scenarios or infrequent updates where GC is infrequent, this may delay conversion and limit the Learned Index's benefits. To address this, we introduce the "Range-query Assisted Conversion" algorithm for proactive conversion.

This proactive conversion is triggered when the LSM's size or level count reaches a predefined threshold, typically when LSM performance begins to degrade. We then perform a full range query across the LSM and Learned Index, merging and sorting key-value pairs from different SST files. Treating the Learned Index as an additional bottom level allows for joint merging and sorting. Using the sorted keys and their Value Log offsets, we construct a new Learned Index while a new LSM and Value Log absorb write operations, ensuring uninterrupted operation.

This approach is lightweight, avoiding key-value pair migration and allowing user-triggered conversion. However, it incurs some overhead for index building without improving space utilization.

\subsubsection{In-memory Bloom Filter}
Our experiments reveal that even with highly skewed workloads (Zipfian distribution), about one-tenth of read queries access the learned index. If the concurrent probing is disabled, our tiered design necessitates an initial LSM check for all queries, as recent keys reside there. This level-by-level LSM traversal, though mitigated by RocksDB's techniques, can accumulate some overhead from false read attempts. To address this, we implement a lightweight in-memory Bloom filter.

We maintain this Bloom filter for keys in the LSM tree, marking representative bits for each new key added to the Memtable. Read requests first consult this filter. If the target key's bits are not fully marked, we skip the LSM and query the Learned Index directly. Otherwise, we query the LSM, accepting a false positive rate under 5\% in our experiments. 

\section{Evaluation}\label{sec:evaluation}
To evaluate our LearnedKV, we compare our design with multiple state-of-the-art KV stores in various test environments. First, in order to show the effectiveness of our tiered design, we construct a micro-benchmark experiment to compare the performance of the LearnedKV with its baseline RocksDB \cite{RocksDB} through YCSB workloads \cite{YCSB} and SOSD datasets \cite{Kipf2019SOSDAB}. Then we broaden our comparison to various situations, including different workload distribution, various dataset size and a range scan experiment. Then, we further include several other state-of-the-art KV stores, Bourbon \cite{Dai2020FromWT}, HashKV \cite{Li2018EnablingEU}, B+-Tree \cite{STX_B+}, and some existing on-storage Learned Indexes into the comparison. 
Finally, we drill down our analysis to our design choice and multiple parameter settings. 


\textbf{Implementation.}
Given that modern workloads typically employ large-sized values \cite{Lu2016WiscKeySK}, we designed our KV store around the KV separation concept. We implemented KV separation for all baselines that did not originally include it. Following prior work \cite{Li2018EnablingEU}, we over-provisioned storage space by 30\% of the key-value pair size, ensuring comparable space allocation across all systems. For fair comparison, we equipped all systems with LearnedKV's GC mechanism, differing only in the indexing approach, while maintaining equivalent memory consumption. We optimized hyper-parameters for each index design and disabled concurrent probing of the Learned Index during LSM searches in LearnedKV.

\textbf{Evaluation Setup.}
We conduct our experiments on a machine with an AMD 64-Core Processor and a SAMSUNG SATA SSD of 447GB on Chameleon testbed \cite{keahey2020lessons}. For computational convenience, unless otherwise specified, the key size and value size are set to 8 bytes and 1016 bytes, respectively, making each key-value pair 1 KB. The pointer/offset size in our testbed is 8 bytes.

\textbf{Workload.} 
In our experiments, we divide the performance test into four phases: P0, P1, P2, and P3. In the first phase P0, we load 10M unique key-value pairs into the KV store. In each subsequent phase (P1, P2, and P3), we perform 10M read or update operations. Given our 1KB key-value pair size, the total data volume processed across these four phases amounts to 40GB (10M operations × 1KB per operation × 4 phases). The number of read or update operations is based on the pre-configured read/write ratio of the workload, with requests following a Zipfian distribution by default. Unless otherwise specified, we collect the read and write throughput from the last phase to mitigate the performance impact of the GC process.

For experiments that do not involve any update or insert operations, we perform a GC once the loading phase (P0) is completed. This is because the loading phase only writes keys into the LSM, and the GC process will not be passively triggered by read-only workloads. Therefore, we manually trigger a GC alongside the conversion from LSM to Learned Index. This procedure is applied uniformly across all schemes, allowing us to highlight the performance differences with and without our Learned Index design more clearly.

In this section, we present the evaluation of LearnedKV by answering the following questions: 
\begin{itemize}
  \item How does the tiered index design benefit the performance? 
  \item How does the LearnedKV compare with other state-of-the-art KV Stores? 
  \item What is the performance of the LearnedKV under various workloads and various key distributions? 
  \item How does the performance get affected by different configurations (e.g. storage over-provision ratio)? 
  \item How much does the performance benefit from our optimizations (e.g. In-memory Bloom Filter)? 
\end{itemize}

\subsection{Overall Performance Comparison}\label{sec:overall performance}
\textbf{Experiment 1: Effect of Learned Index.} 
We evaluate the impact of the Learned Index using a read-write balanced (50\% read and 50\% write) YCSB workload with a Zipfian distribution (s=0.99). The experiment involves loading 10M key-value pairs followed by three operational phases, each consisting of 5M read and 5M update operations.

\begin{figure}[htbp]
  \centering
  \includegraphics[width=\linewidth]{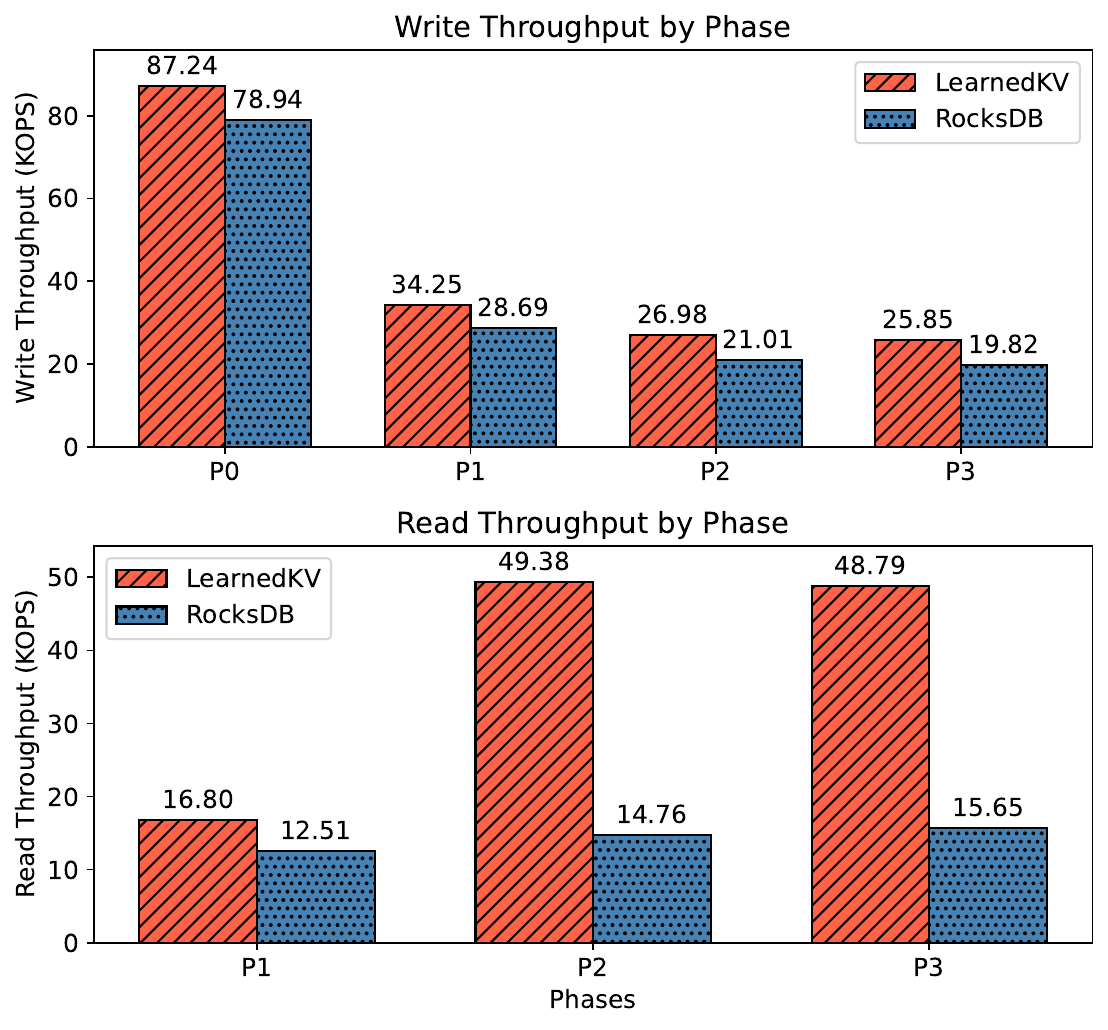}
  \caption{\textbf{Throughput Comparison between LearnedKV and RocksDB.} \textit{P0 is load phase; P1,P2,P3 consist of read and update. The learned Index is built in the middle of P1. GC happens in P1,P2,P3)}}
  \label{fig:phase_throughput}
\end{figure}

Figure \ref{fig:phase_throughput} shows the overall throughput comparison among LearnedKV, and RocksDB+ over different phases. "RocksDB+" is the modified implementation of RocksDB to which we adapted our key-value separation and GC mechanism. The only difference between LearnedKV and RocksDB+ is that LearnedKV has a tiered layer of Learned Index while RocksDB dose not. In this way, we can clearly see the effectiveness of our key design concept. 

From the results, we can see that, in Phase 0 (loading), both systems perform similarly as the Learned Index is not yet built. During P1, LearnedKV begins to show advantages, because the Learned Index construction is triggered in the middle of P1. The performance gap gets wider in P2 and P3, where LearnedKV fully leverages its tiered index structure. LearnedKV outperforms RocksDB+ by up to 1.30x in write throughput and 3.35x in read throughput during these phases, clearly demonstrating the effectiveness of the Learned Index.

Table \ref{tab:size_comparison} illustrates the storage space usage at the end of P3. While Key-Value Space consumption remains constant due to identical workloads and GC policies, LearnedKV significantly reduces the indexing size (2.98x). This improvement is largely contributed from the space amplification within the LSM. 

\begin{table}[htbp]
    \centering
    \begin{tabular}{lcc}
        \toprule
        & \textbf{LearnedKV} & \textbf{RocksDB} \\
        \midrule
        \textbf{Key-Ptr Space} & & \\
        LSM & 4.2MB & 242MB \\
        key\_array & 77MB & - \\
        model & 96KB & - \\
        \textbf{Total} & \textbf{81.3MB} & \textbf{242MB} \\
        \midrule
        \textbf{Key-Value Space} & & \\
        vlog & 9.6GB & 9.6GB \\
        \bottomrule
    \end{tabular}
    \caption{\textbf{Comparison of storage sizes for LearnedKV and RocksDB.}}
    \label{tab:size_comparison}
\end{table}

\textbf{Experiment 2: Read-Write Ratio.}
To comprehensively assess the impact of our design, we extend Experiment 1 to evaluate performance across different read/write ratios. We examine four scenarios of different read:write ratios: read-heavy (7:3), read-write-balanced (5:5), write-heavy (3:7), and write-only workloads. The experiment involves loading key-value pairs followed by three operational phases with the specified ratios, focusing on updates to existing keys. We analyze the last phase to minimize GC process impact and exclude read-only workloads as they don't trigger GC,  which, in such cases, cannot show the key design of LearnedKV. 

\begin{figure}[htbp]
  \centering
  \includegraphics[width=\linewidth]{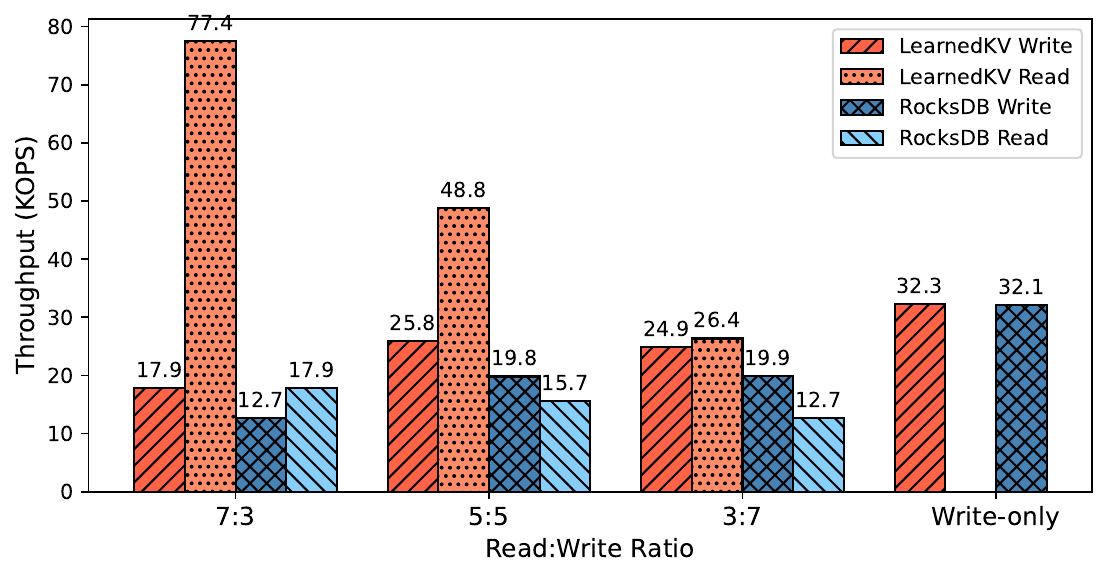}
  \caption{\textbf{Performance of LearnedKV vs. RocksDB for Different Read: Write Ratios}}
  \label{fig:readwrite_ratio}
\end{figure}


Figure \ref{fig:readwrite_ratio} demonstrates that LearnedKV can consistently improves operation throughput across all scenarios, with gains ranging up to 1.41x in write throughput and 4.32x in read throughput. This stability in read and write performance enhancement across various workloads underscores the robustness of LearnedKV's design in handling diverse operational patterns.

\textbf{Experiment 3: SOSD dataset.}
We also evaluate our design on real-world complex datasets collected by SOSD \cite{Kipf2019SOSDAB}, which is designed for Learned Index. We used four of their datasets: Amazon book sale popularity (books) \cite{books}, unsampled Facebook user IDs (fb) \cite{Sandt2019EfficientlySI}, uniformly sampled OpenStreetMap locations (osm) \cite{Pandey2018HowGA}, and Wikipedia article edit timestamps (wiki) \cite{wiki}. As shown in Table \ref{tab:key-space-comparison}, LearnedKV demonstrates consistent performance advantages across all datasets. For write operations, LearnedKV achieves improvements ranging from 1.18x to 1.25x over RocksDB+. The read performance improvements are even more substantial, with LearnedKV outperforming RocksDB+ by 2.69x to 3.21x. Most notably, on the Facebook user ID dataset, LearnedKV achieves a 3.21x read throughput improvement while maintaining 1.24x better write performance, demonstrating the system's ability to handle real-world data distributions effectively.

\begin{table}[htbp]
\centering
\resizebox{\linewidth}{!}{%
\begin{tabular}{|c|c|c|c|c|}
\hline
\textbf{Dataset} & \multicolumn{2}{c|}{\textbf{Write (KOPS)}} & \multicolumn{2}{c|}{\textbf{Read (KOPS)}} \\
\cline{2-5}
\textbf{Size} & \textbf{LearnedKV} & \textbf{RocksDB+} & \textbf{LearnedKV} & \textbf{RocksDB+} \\
\hline
books & 24.31 \textbf{(1.18x)} & 20.53 & 50.81 \textbf{(2.72x)} & 18.65 \\
fb & 25.34 \textbf{(1.24x)} & 20.36 & 54.38 \textbf{(3.21x)} & 16.92 \\
osm & 24.88 \textbf{(1.20x)} & 20.70 & 52.52 \textbf{(3.17x)} & 16.55 \\
wiki & 43.83 \textbf{(1.25x)} & 35.18 & 54.29 \textbf{(2.69x)} & 20.21 \\
\hline
\end{tabular}%
}
\caption{\textbf{Performance comparison of LearnedKV and RocksDB+ under different key space distributions.} \textit{KOPS: Thousand Operations Per Second}}
\label{tab:key-space-comparison}
\end{table}

\textbf{Experiment 4: Workload Distribution.}
Table \ref{tab:workload-comparison} demonstrates LearnedKV's consistent superiority over RocksDB+ across various workload distributions. For write operations, LearnedKV shows improvements of 1.04x to 1.30x. Read performance advantages are even more significant, especially in skewed distributions. Under Zipfian distributions (parameters 0.99 and 0.9), LearnedKV achieves impressive 3.12x and 2.95x read throughput improvements respectively. Even with uniform distribution, LearnedKV maintains a 2.51x read performance advantage. These results highlight LearnedKV's robust performance across diverse workload patterns, particularly benefiting read-heavy workloads in skewed distributions common in real-world scenarios.

\begin{table}[htbp]
\centering
\resizebox{\linewidth}{!}{%
\begin{tabular}{|c|c|c|c|c|}
\hline
\textbf{Dataset} & \multicolumn{2}{c|}{\textbf{Write (KOPS)}} & \multicolumn{2}{c|}{\textbf{Read (KOPS)}} \\
\cline{2-5}
\textbf{Size} & \textbf{LearnedKV} & \textbf{RocksDB+} & \textbf{LearnedKV} & \textbf{RocksDB+} \\
\hline
Zipfian(0.99) & 25.85 \textbf{(1.30x)} & 19.82 & 48.79 \textbf{(3.12x)} & 15.65 \\
Zipfian(0.9) & 40.29 \textbf{(1.04x)} & 38.58 & 50.79 \textbf{(2.95x)} & 17.24 \\
Uniorm & 35.36 \textbf{(1.08x)} & 32.80 & 11.14 \textbf{(2.51x)} & 4.43 \\
\hline
\end{tabular}%
}
\caption{\textbf{Performance comparison of LearnedKV and RocksDB+ under different workload distributions.} \textit{KOPS: Thousand Operations Per Second}}
\label{tab:workload-comparison}
\end{table}




\textbf{Experiment 5: Dataset size.}
Table \ref{tab:dataset-size-comparison} shows LearnedKV's performance across dataset sizes from 500K to 10M key-value pairs. LearnedKV consistently outperforms RocksDB+, with write improvements of 1.08x to 1.35x. Read performance advantages grow significantly with dataset size, from 1.27x at 500K to 3.12x at 10M entries. Even with 10M entries, LearnedKV maintains substantial advantages in both writes (25.85 KOPS, 1.30x improvement) and reads (48.79 KOPS, 3.12x improvement), demonstrating effective scalability particularly for read operations.

\begin{table}[htbp]
\centering
\resizebox{\linewidth}{!}{%
\begin{tabular}{|c|c|c|c|c|}
\hline
\textbf{Dataset} & \multicolumn{2}{c|}{\textbf{Write (KOPS)}} & \multicolumn{2}{c|}{\textbf{Read (KOPS)}} \\
\cline{2-5}
\textbf{Size} & \textbf{LearnedKV} & \textbf{RocksDB+} & \textbf{LearnedKV} & \textbf{RocksDB+} \\
\hline
500K & 54.96 \textbf{(1.08x)} & 50.56 & 97.09 \textbf{(1.27x)} & 76.48 \\
1M & 38.00 \textbf{(1.14x)} & 33.40 & 85.54 \textbf{(1.90x)} & 44.94 \\
3M & 31.74 \textbf{(1.35x)} & 23.57 & 62.38 \textbf{(2.95x)} & 21.15 \\
6M & 26.92 \textbf{(1.24x)} & 21.69 & 51.61 \textbf{(2.90x)} & 17.80 \\
10M & 25.85 \textbf{(1.30x)} & 19.82 & 48.79 \textbf{(3.12x)} & 15.65 \\
30M & 22.24 \textbf{(1.12x)} & 19.77 & 23.01 \textbf{(1.87x)} & 12.33 \\
\hline
\end{tabular}%
}
\caption{\textbf{Performance comparison of LearnedKV and RocksDB+ with different dataset sizes.} \textit{KOPS: Thousand Operations Per Second}}
\label{tab:dataset-size-comparison}
\end{table}

\textbf{Experiment 6: Range Scan.}
We also compare our range scan performance with the state-of-the-art RocksDB. Similar to previous experiments, we first load 1,000,000 key-value pairs into the KV store and perform 1,000 scan operations over the database. For the range scan, we did not pre-set the number of key-value pairs to be involved, as these numbers are impractical to pre-determine in real-world scenarios. Instead, we configured a \textit{ScanRange}, and all KV stores were used to read all the key-value pairs within this range. To ensure a fair comparison, we modified the string keys used in RocksDB so that it could maintain the same order as integers. This ensured that the keys involved in each competitor were identical. Based on the experiment logs, each scan request in this experiment reads about 500KB of key-value pairs on average.

\begin{figure}[htbp]
  \centering
  \includegraphics[width=\linewidth]{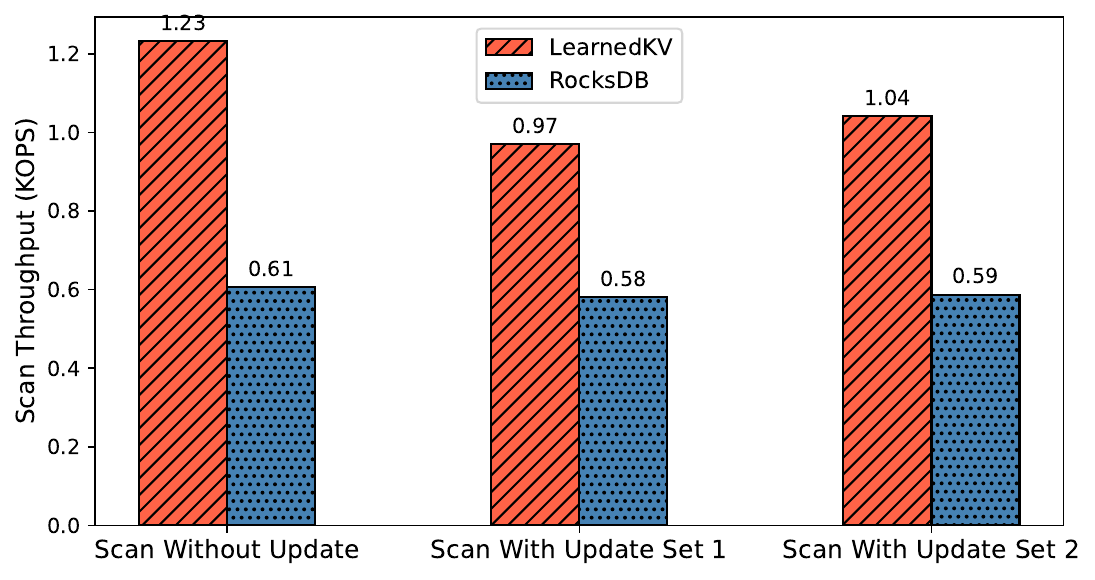}
  \caption{\textbf{Performance Comparison of Range Scan}}
  \label{fig:range_scan}
\end{figure}

As shown in figure \ref{fig:range_scan}, our setup consists of three parts, using the same loading phase but with different workloads. "Scan without update" workloads consist solely of scan operations. In LearnedKV, after the GC process at the end of the loading phase, all valid key-value pairs are re-grouped and migrated to the Learned Index. Because there are no insert or update requests afterward, the LSM will not absorb any new keys, making the Learned Index the only active component in this tiered storage. Thus, this comparison effectively measures the performance difference between the Learned Index and RocksDB. In this scenario, LearnedKV achieves up to 2.02x greater performance in range scans compared to RocksDB. However, such scenarios are rare. Therefore, we also conducted experiments on "Scan with Update" workloads. We included two sets of such workloads: "Set 1" contains 0.5M updates before the scan requests, while "Set 2" includes 1M update requests before the scan. These workloads make the LSM absorb some keys after the GC process, requiring LearnedKV to scan both the LSM and the Learned Index to obtain correct results. As expected, there is a 21\% performance drop compared to the pure scan workload, but it still outperforms the state-of-the-art by 1.76x.


\textbf{Experiment 7: Other KV Stores.} 
We evaluated LearnedKV against several KV stores optimized for on-storage performance: Bourbon \cite{Dai2020FromWT}, HashKV \cite{Li2018EnablingEU}, disk-resident B+-Tree \cite{STX_B+}, and hybrid learned indexes \cite{Zhang2024MakingIL}. Since most of the works are adapted from LevelDB, for a fair comparison, we also implemented a LevelDB-based version of LearnedKV and adapted B+-Tree and hybrid indexes with key-value separation and garbage collection mechanisms similar to LearnedKV. The B+-Tree was configured with leaf nodes and two levels of inner nodes on storage to match memory consumption. For hybrid indexes, we used "Hybrid\_PGM\_Disk" and "Hybrid\_Leco\_Disk" that are proposed as "all-in-one" hybrid index from \cite{Zhang2024MakingIL}. We also fine-tuned all the KV-stores such that all memory consumption will be similar.

\begin{figure}[htbp]
  \centering
  \includegraphics[width=\linewidth]{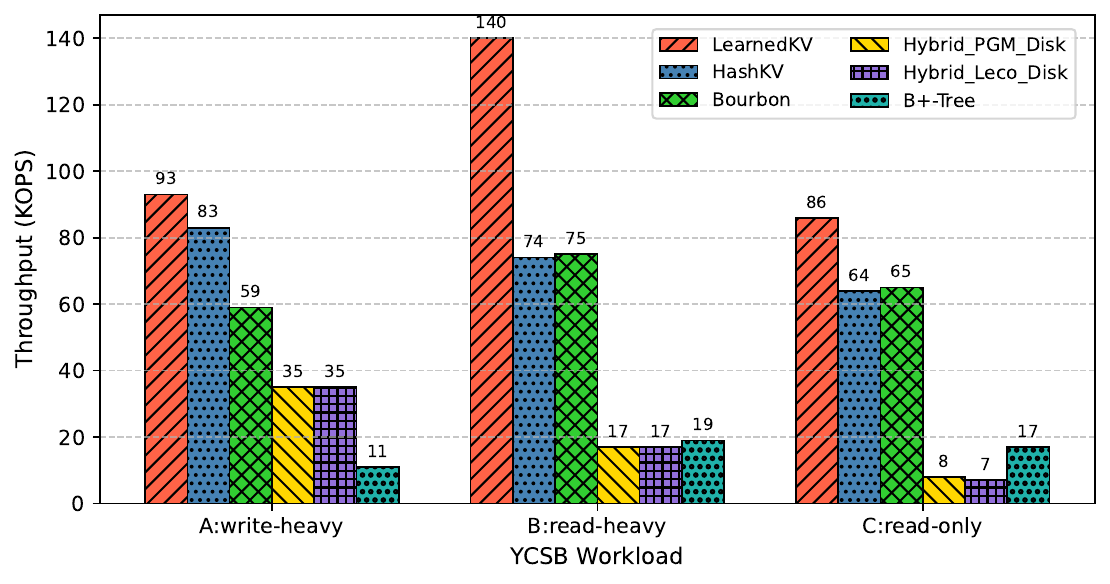}
  \caption{\textbf{KV Store Throughput Comparison for YCSB Workloads.} \textit{Three YCSB workloads: A (write-heavy, w:50\%, r:50\%), B (read-heavy, w:5\%, r:95\%), and C (read-only).}}
  \label{fig:kvstore}
\end{figure}

Figure \ref{fig:kvstore} compares throughput across these KV stores under YCSB workloads A (write-heavy), B (read-heavy), and C (read-only). In this experiment, we use the benchmark with 1M entries because when the data size is larger, some existing solutions will become extremely slow (> 2 hours for each datapoint). LearnedKV consistently outperforms all systems across workload types. In the write-heavy workload, LearnedKV surpasses HashKV by 12\% and Bourbon by 58\%. This performance advantage becomes more pronounced in read-heavy workloads, with LearnedKV outperforming HashKV by 89\% and Bourbon by 87\%. For the read-only scenario, LearnedKV maintains a 34\% and 32\% lead over HashKV and Bourbon, respectively. To be noticed that, one of the important reason of the extreme well performance of LearnedKV on read-heavy workloads, based on our analysis, is that the Learned Index help LSM reduce its size so that many of the read requests can be efficiently processed by both LSM and Learned Index.

LearnedKV's advantages are even more pronounced against on-storage learned indexes and B+-Trees. In write-heavy workloads, it outperforms them by 2.66x. This performance gap widens dramatically in read-intensive workloads, with LearnedKV achieving up to 8.24x higher throughput in read-heavy scenarios.


\begin{table}[htbp]
\centering
\resizebox{\linewidth}{!}{%
\begin{tabular}{|c|c|c|c|}
\hline
\textbf{YCSB} & \multicolumn{3}{c|}{\textbf{Throughput (KOPS)}}\\
\cline{2-4}
\textbf{workload} & \textbf{LearnedKV} & \textbf{Hybrid\_PGM\_Disk}  & \textbf{Hybrid\_Leco\_Disk}\\
\hline
A & 48.52 & 34.62 & 35.37 \\
B & 90.78 & 17.16 & 16.96 \\
C & 60.80 & 7.66 & 7.43 \\
\hline
\end{tabular}%
}
\caption{\textbf{Performance comparison of LearnedKV and on storage Learned Index with direct I/O} \textit{KOPS: Thousand Operations Per Second}}
\label{tab:direct_io}
\end{table}

We also evaluated LearnedKV against state-of-the-art on-storage learned indexes from \cite{Zhang2024MakingIL} under direct I/O, using RocksDB as our base LSM implementation. (Direct I/O is not supported in LevelDB and its variants mentioned above.) As shown in Table \ref{tab:direct_io}, LearnedKV can still significantly outperform both Hybrid\_PGM\_Disk and Hybrid\_Leco\_Disk across all tested YCSB workloads. Most notably, for read-heavy workload B, LearnedKV achieves 5.3x higher throughput, while for read-only workload C, the performance gap widens to 7.9x.

\subsection{Drill-down Analysis}
To fully investigate what is happening inside the LearnedKV, we further developed some drill-down analysis on its internal behavior.

\begin{figure}[htbp]
  \centering
  \includegraphics[width=\linewidth]{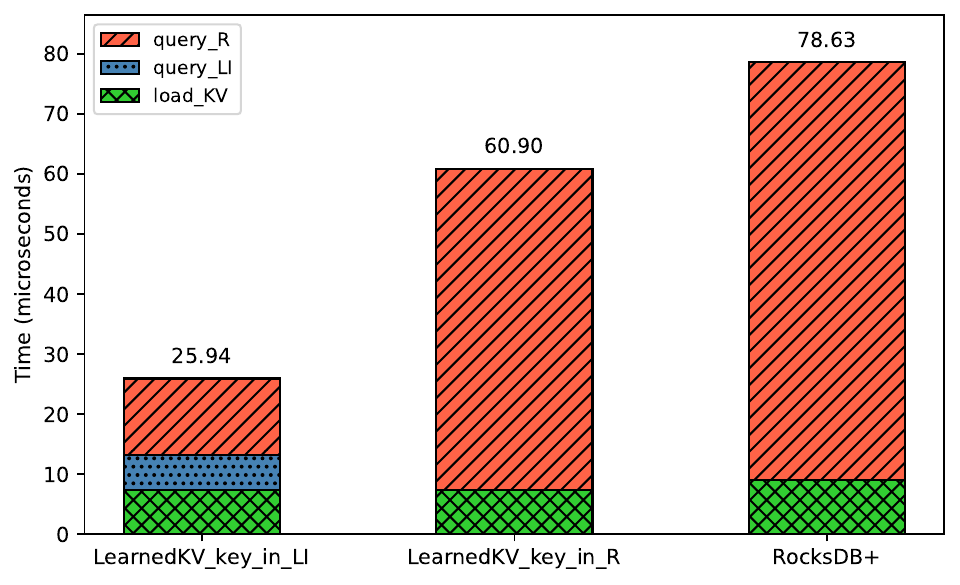}
  \caption{\textbf{Time breakdown of read operations over LearnedKV and RocksDB+}}
  \label{fig:time_breakdown}
\end{figure}

\textbf{Experiment 8: Latency \& Time breakdown.} Table \ref{tab:time_breakdown} and Figure \ref{fig:time_breakdown} provide a detailed time breakdown of LearnedKV read operations on Phase 1. Each operation consists of three components: RocksDB querying, Learned Index querying, and KV pair loading from the vlog.

Among approximately 5M read operations, 12.6\% are processed by the Learned Index with an average time of 25.94 \si{\micro\second}, either through LSM read failures or by skipping LSM entirely. Despite using an in-memory Bloom Filter, over 95\% of read requests still access RocksDB, a result of the skewed workload favoring frequently updated keys.

LearnedKV handles read queries in two scenarios: (1) "LearnedKV\_key\_in\_R", where LSM successfully locates and returns the key, and (2) "LearnedKV\_key\_in\_LI", where either LSM fails to find the key or the Bloom Filter indicates the key's absence in LSM, triggering a Learned Index query. Figure \ref{fig:time_breakdown} shows that Learned Index queries achieve 67\% lower latency compared to the baseline, despite requiring both indexes, due to efficient LSM failure detection and fast Learned Index reads. Even for LSM-resident keys, LearnedKV delivers 23\% lower latency, benefiting from its reduced LSM size compared to RocksDB+.

\begin{table}[htbp]
\centering
\begin{tabular}{l|r}
\hline
\textbf{Operation} & \textbf{Number} \\
\hline
Write & 5,001,245 \\
Read & 4,998,755 \\
\hline
\multicolumn{2}{l}{\textbf{Read Operation Breakdown}} \\
\hline
Read through LI & 538,347 \\
Read through RD & 4,763,353 \\
Load from vlog & 4,998,755 \\
\hline
\end{tabular}
\caption{\textbf{Summary of operations and their counts.} \textit{LI: Learned Index; RD: RocksDB}}
\label{tab:time_breakdown}
\end{table}

\textbf{Experiment 9: Over-provisioning ratio.}
Table \ref{tab:over-provisioning} compares LearnedKV and RocksDB+ performance across various over-provisioning ratios. LearnedKV consistently outperforms RocksDB+ in both write (1.14x to 1.29x improvement) and read operations (1.39x to 1.65x improvement). Write performance for both systems improves with higher over-provisioning, with LearnedKV peaking at 59.57 KOPS at 50\% over-provisioning. Read performance in LearnedKV shows more variation (69.77 to 76.34 KOPS) compared to RocksDB+'s stability. These results demonstrate LearnedKV's superior performance, particularly in read operations, while highlighting the trade-offs in over-provisioning ratio selection, as it differentially affects read and write performance.

\begin{table}[htbp]
\centering
\resizebox{\linewidth}{!}{%
\begin{tabular}{|c|c|c|c|c|}
\hline
\textbf{Over-} & \multicolumn{2}{c|}{\textbf{Write (KOPS)}} & \multicolumn{2}{c|}{\textbf{Read (KOPS)}} \\
\cline{2-5}
\textbf{prov. (\%)} & \textbf{LearnedKV} & \textbf{RocksDB+} & \textbf{LearnedKV} & \textbf{RocksDB+} \\
\hline
10 & 11.99 \textbf{(1.15x)} & 10.41 & 37.23 \textbf{(3.01x)} & 12,37 \\
20 & 26.84 \textbf{(1.42x)} & 18.94 & 46.36 \textbf{(3.06x)} & 15.16 \\
30 & 24.71 \textbf{(1.29x)} & 19.19 & 49.48 \textbf{(3.11x)} & 15.91 \\
40 & 36.60 \textbf{(1.22x)} & 29.88 & 46.95 \textbf{(2.95x)} & 15.90 \\
50 & 36.35 \textbf{(1.30x)} & 27.86 & 40.18 \textbf{(2.58x)} & 15.59 \\
\hline
\end{tabular}%
}
\caption{\textbf{Performance comparison of LearnedKV and RocksDB+ with different over-provisioning ratios} \textit{KOPS: Thousand Operations Per Second}}
\label{tab:over-provisioning}
\end{table}

\begin{figure}[htbp]
  \centering
  \includegraphics[width=\linewidth]{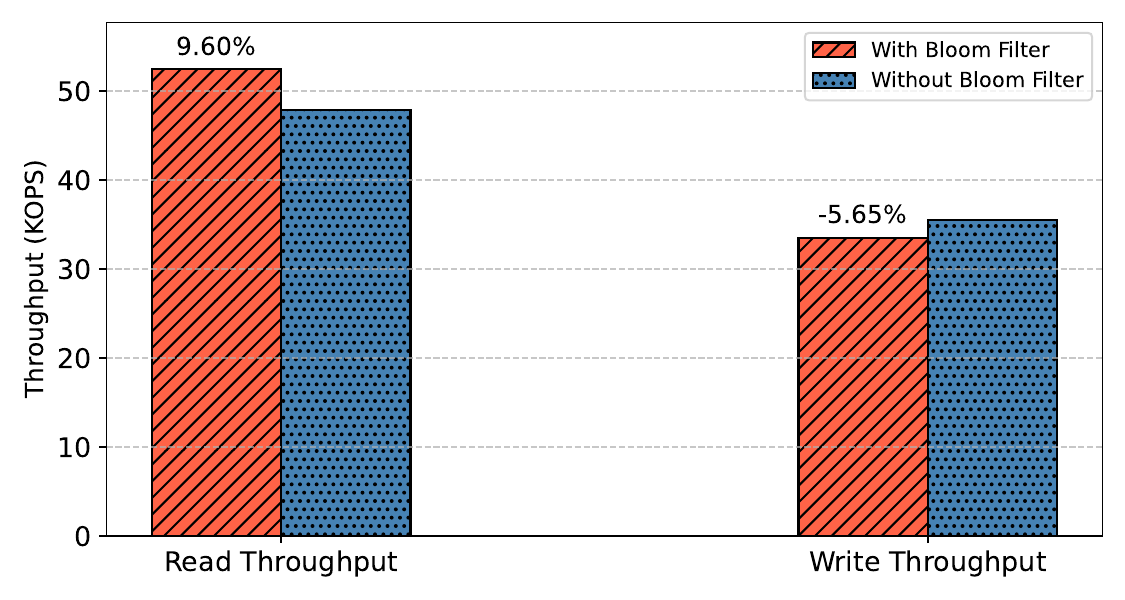}
  \caption{\textbf{Comparison between throughputs with and without Bloom Filter}}
  \label{fig:bloom_filter}
\end{figure}

\textbf{Experiment 10: In-memory Bloom Filter.}
Figure \ref{fig:bloom_filter} shows the performance comparison related to the in-memory Bloom Filter. After applying the Bloom Filter, write throughput decreases by 5.65\% due to maintenance overhead for new key insertions. The read throughput improves by 9.60\%, providing a modest performance gain. This limited improvement is not surprising since, according to our previous experiments \ref{tab:time_breakdown}, the Bloom Filter only affects about 0.235M among 5M operations by directing them directly to the Learned Index. However, this presents an interesting trade-off: a small write performance penalty for potentially significant read gains in read-heavy workloads, making the Bloom Filter a valuable optimization option depending on workload characteristics.

\subsection{Performance in Memory}
\begin{figure}[ht]
  \centering
  \includegraphics[width=\linewidth]{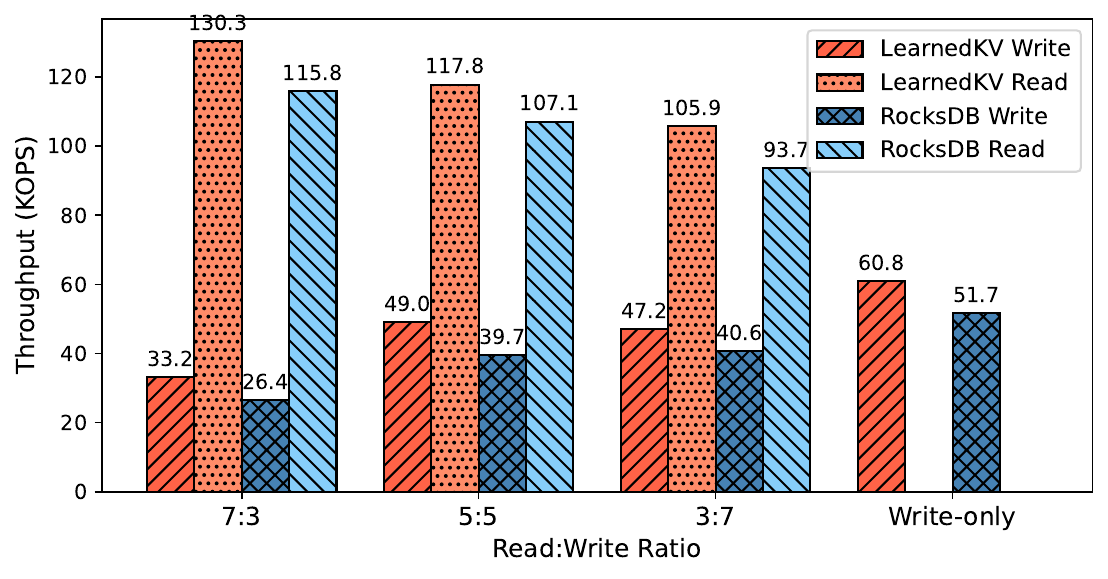}
  \caption{\textbf{Performance of LearnedKV vs. RocksDB for Different Read-Write Ratios in Memory}}
  \label{fig:readwrite_ratio_memory}
\end{figure}

To evaluate the in-memory performance characteristics of both systems, we conducted experiments with datasets that fit entirely in DRAM. Figure \ref{fig:readwrite_ratio_memory} illustrates the throughput comparison between LearnedKV and RocksDB across various read:write ratios. LearnedKV consistently outperforms RocksDB by up to 1.26x. This demonstrates that LearnedKV's architectural benefits extend beyond storage-bound scenarios to in-memory operations, where the tiered index design effectively reduces computational overhead and improves cache efficiency.

\subsection{Performance on HDD}

\begin{figure}[ht]
  \centering
  \includegraphics[width=\linewidth]{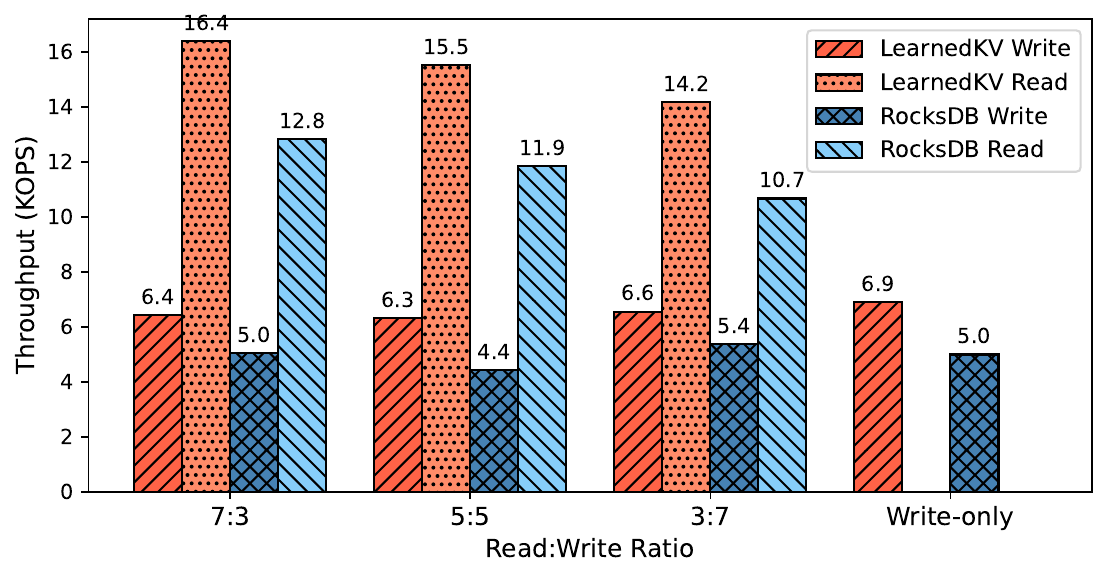}
  \caption{\textbf{Performance of LearnedKV vs. RocksDB for Different Read-Write Ratios on HDD}}
  \label{fig:readwrite_ratio_HDD}
\end{figure}

To evaluate LearnedKV's adaptability across diverse storage environments, we extended our experiments to HDD-based systems using a 12-core Intel Xeon E5-2620 v3 2.40GHz CPU with a 931GB Seagate Constellation ES.3 HDD, and we used 1M-key datasets due to the slow performance of HDD. Figure \ref{fig:readwrite_ratio_HDD} compares LearnedKV's performance against RocksDB for various read-write ratios on this HDD setup. LearnedKV demonstrates more pronounced advantages on HDDs compared to SSDs, consistently outperforming RocksDB across all scenarios with throughput improvements of 1.22x to 1.43x. Notably, LearnedKV shows significant gains in write-heavy and write-only workloads, which are typical bottlenecks for HDD-based systems. The substantial write throughput improvement (up to 1.43x) highlights LearnedKV's effectiveness in mitigating mechanical storage limitations. These results underscore LearnedKV's versatility and its capacity to optimize performance across both SSD and HDD infrastructures, effectively addressing the challenges posed by mechanical storage systems.

\section{Future Work \& Conclusion}



The choice of the Learned Index model presents an important part of future research. While our implementation uses piece-wise linear representation for its simplicity and efficiency, alternative learned index structures could potentially offer different performance trade-offs for on-storage indexing. This represents a rich area for future exploration.

In this paper, we introduced LearnedKV, a novel approach to key-value store design through its tiered architecture that fully decouples the LSM tree from the Learned Index. By having the LSM tree handle write operations while the Learned Index accelerates reads, the system achieves superior performance compared to traditional approaches that treat learned indexes as auxiliary components. Our non-blocking conversion mechanism efficiently migrates data during garbage collection, enabling smooth operation without sacrificing performance. This design significantly reduces LSM size, which leads to improvements in both read and write operations.

Extensive experimental evaluation demonstrates that LearnedKV consistently outperforms state-of-the-art LSM-based solutions across diverse workloads and environments, achieving up to 4.32x faster reads and 1.43x faster writes. The system maintains these performance advantages across different data distributions, workload patterns, and storage media, confirming the robustness and broad applicability of our approach.


{\footnotesize \bibliographystyle{acm}
\bibliography{usenix}}

\end{document}